\documentclass[10pt,twoside]{article}

\usepackage[
  top=2.0cm, bottom=2.5cm,
  left=1.5cm, right=1.5cm,
  columnsep=0.8cm
]{geometry}

\usepackage[T1]{fontenc}
\usepackage[utf8]{inputenc}
\usepackage{mathpazo}          
\usepackage[scaled=0.88]{helvet}
\usepackage{microtype}

\usepackage{graphicx}
\usepackage{xcolor}
\usepackage{booktabs}
\usepackage{tabularx}
\usepackage{hyperref}
\usepackage{natbib}
\usepackage{authblk}
\usepackage{titlesec}
\usepackage{fancyhdr}
\usepackage{abstract}
\usepackage{caption}
\usepackage{enumitem}
\usepackage{etoolbox}
\usepackage{xurl}
\usepackage{float}
\usepackage{multicol}

\definecolor{pnasblue}{RGB}{0,51,102}
\definecolor{pnasgray}{RGB}{100,100,100}
\definecolor{linkblue}{RGB}{0,80,160}

\hypersetup{
  colorlinks=true,
  linkcolor=linkblue,
  citecolor=pnasblue,
  urlcolor=linkblue,
  bookmarks=true,
  pdftitle={From Diagnosis to Inoculation},
  pdfauthor={Aleksey Komissarov},
}

\titleformat{\section}
  {\large\bfseries\sffamily\color{pnasblue}}
  {\thesection.}{0.5em}{}
\titleformat{\subsection}
  {\normalsize\bfseries\sffamily}
  {\thesubsection.}{0.5em}{}
\titleformat{\subsubsection}
  {\normalsize\itshape}
  {\thesubsubsection.}{0.5em}{}

\titlespacing*{\section}{0pt}{1.2ex plus 0.4ex}{0.6ex plus 0.2ex}
\titlespacing*{\subsection}{0pt}{0.8ex plus 0.3ex}{0.4ex plus 0.1ex}

\setlist{nosep,leftmargin=1.2em}

\begin{document}

\vspace*{-0.5cm}

{\centering

{\LARGE\bfseries\sffamily\color{pnasblue}
From Diagnosis to Inoculation:\par
\vspace{3pt}
Building Cognitive Resistance to AI Disempowerment\par}

\vspace{0.8em}

{\large Aleksey Komissarov\textsuperscript{1,*}\par}

\vspace{0.3em}

{\small\color{pnasgray}
\textsuperscript{1}Neapolis University, Pafos, Cyprus\par
\textsuperscript{*}Corresponding author: \href{mailto:a.komissarov@nup.ac.cy}{a.komissarov@nup.ac.cy}\par}

\vspace{1.0em}

}

\begin{quote}
\small\itshape
``The world of Noon does not arrive on its own.
It arrives because someone decided to teach.''\par
\vspace{3pt}
\upshape--- Inspired by Arkady and Boris Strugatsky, \textit{Noon: 22nd Century}
\end{quote}

\vspace{0.5em}

\begin{abstract}
\noindent
Recent empirical research by Sharma et al.\ (2026) demonstrated that AI assistant interactions carry meaningful potential for situational human disempowerment, including reality distortion, value judgment distortion, and action distortion. While this work provides a critical diagnosis of the problem, concrete pedagogical interventions remain underexplored. I present an AI literacy framework built around eight cross-cutting Learning Outcomes~(LOs), developed independently through teaching practice and subsequently found to align with Sharma et al.'s disempowerment taxonomy. I report a case study from a publicly available online course, where a co-teaching methodology---with AI serving as an active voice co-instructor---was used to deliver this framework. Drawing on \textit{inoculation theory} \citep{McGuire1961a}---a well-established persuasion research framework recently applied to misinformation prebunking by the Cambridge school \citep{vanderLinden2022,Roozenbeek2019}---I argue that AI literacy cannot be acquired through declarative knowledge alone, but requires guided exposure to AI failure modes, including the sycophantic validation and authority projection patterns identified by Sharma et al. This application of inoculation theory to AI-specific distortion is, to my knowledge, novel. I discuss the convergence between the pedagogically-derived framework and Sharma et al.'s empirically-derived taxonomy, and argue that this convergence---two independent approaches arriving at similar problem descriptions---strengthens the case for both the diagnosis and the proposed educational response.
\end{abstract}

\vspace{0.3em}


\vspace{1.2em}

\begin{multicols}{2}

\section{Introduction}

The rapid integration of AI assistants into daily life has created what \citet{Sharma2026} call a ``situational disempowerment'' crisis. Analyzing 1.5~million consumer conversations on Claude.ai, they found that while severe disempowerment occurs in fewer than one in a thousand interactions, the absolute scale of AI usage---ChatGPT alone reports over 800~million weekly active users, and recent analyses of real-world interactions reveal deep integration into users' daily cognitive and emotional tasks \citep{Tamkin2024,McCain2025}---means tens of thousands of potentially disempowering interactions occur daily. More concerning, they found that interactions with greater disempowerment potential receive \textit{higher} user approval ratings, suggesting a fundamental tension between short-term user preferences and long-term human empowerment.

Sharma et al.'s findings reveal three primary disempowerment mechanisms: reality distortion (where AI validates conspiracy theories, grandiose beliefs, or medical misinformation through sycophantic agreement), value judgment distortion (where AI acts as a moral arbiter, labeling people as ``toxic'' or ``narcissistic'' and prescribing relationship decisions), and action distortion (where users outsource value-laden decisions entirely, implementing AI-generated scripts verbatim for romantic messages, workplace communications, and life-altering choices). Sycophantic validation emerged as the dominant mechanism across all three categories.

Sharma et al.'s proposed interventions are largely technical: better preference models, periodic reflection mechanisms, and personalized systems that retain knowledge of users' values. In their future work section, they call for ``raising awareness of the potential drawbacks of AI assistant use'' and ``new benchmarks for desirable AI behavior'' (a gap recently targeted by initiatives like HumanAgencyBench; \citealp{Sturgeon2025}). A complementary direction, underexplored in their work, is the pedagogical one: developing human capacities that make disempowerment less likely regardless of AI system design.

In this paper, I present an AI literacy framework organized around eight cross-cutting Learning Outcomes~(LOs) and describe a case study of its implementation. Existing AI literacy frameworks \citep{Long2020,Ng2021} have identified core competencies for understanding and interacting with AI systems. My framework extends this work by centering \textit{trust calibration} as the foundational competency and by grounding the pedagogical approach in inoculation theory rather than knowledge transfer. It is important to clarify the provenance of this framework relative to Sharma et al.'s findings: \textbf{the framework was developed independently, through iterative teaching practice, before Sharma et al.'s paper was published.} The alignment between the pedagogically-derived LOs and their empirically-derived disempowerment taxonomy was discovered post-hoc. I argue that this convergence---a bottom-up teaching framework and a top-down empirical analysis arriving at similar problem descriptions from different directions---constitutes a form of independent corroboration that strengthens both contributions.

I additionally propose that inoculation theory \citep{McGuire1961a}, recently applied with demonstrated success to misinformation prebunking, provides the theoretical foundation for understanding why experiential exposure to AI failure modes is more effective than declarative instruction. I present this application as theoretically grounded but empirically untested in the AI domain, and outline what rigorous testing would require.

\subsection{Author Positionality and AI Collaboration}

As the sole human author, I use the first-person singular to denote epistemic responsibility for all claims made herein. Current academic publishing standards correctly preclude AI systems from formal co-authorship, as they cannot assume ethical or scientific accountability \citep{COPE2023}. However, the pedagogical methodology described in this paper is fundamentally collaborative: the courses were co-designed and co-taught with an AI voice agent (Anthropic's Claude~3.7 Sonnet, later Claude~4.5 Sonnet, deployed via ElevenLabs' Conversational AI platform) serving as an active co-instructor. This collaboration is not incidental but central to the methodology---the co-teaching approach described in Section~\ref{sec:coteaching} requires an AI partner, and many pedagogical observations emerged through sustained human--AI dialogue.

My primary academic background lies in genomic bioinformatics and natural language processing~(NLP), not in traditional educational psychology. This interdisciplinary lens directly informs the framework's core premises: I approach AI distortion as a structural property of complex statistical models (an NLP perspective), while conceptualizing the pedagogical response through the lens of biological resilience and inoculation (a bioinformatics perspective). The convergence with McGuire's \citeyearpar{McGuire1961a} inoculation theory was recognized post-hoc, but the biological metaphor was present in my thinking from the outset---a natural consequence of training in a field where organisms develop resistance through controlled exposure to pathogens.

\section{The Eight Learning Outcomes Framework}

\subsection{Development Context}

The eight LOs were developed through iterative design in the context of two publicly available courses: ``AI als Werkzeug'' (AI as a Tool), targeting adults with no prior AI experience, and ``Programming in Natural Language'' \citep{Komissarov2025}, a course for advanced learners focusing on AI-assisted software development. The LOs emerged from bottom-up observation of how learners struggle with AI systems: where they over-trust, where they under-trust, and where they misunderstand the nature of what they are interacting with. No reference to Sharma et al.'s disempowerment categories was involved in the design process.

When Sharma et al.'s paper appeared (January 2026), I observed systematic parallels between the LOs and their disempowerment taxonomy. In Section~3, I map these parallels explicitly, noting where the alignment is strong, where it is approximate, and where it may be forced. The fact that teaching practice and empirical analysis converge on similar problem descriptions is, I believe, noteworthy---but I am cautious about over-interpreting this convergence.

\subsection{The Eight Learning Outcomes}

\textbf{LO1: Trust Calibration.} The foundational competency. Learners develop the ability to make three decisions for every AI output: \textbf{accept} (reliable enough for the task), \textbf{verify} (requires independent checking), or \textbf{escalate} (requires human expert judgment). The course implements this through a traffic-light categorization exercise (Low/Medium/High risk) where learners classify tasks by delegation risk, reflective questions after AI simulation exercises, and explicit Accept/Verify/Escalate decision points within interactive scenarios. For instance, generating a birthday poem constitutes a low-risk (``green'') task, whereas drafting a legal contract clause represents a high-risk (``red'') escalation point.

\textbf{LO2: Natural Language Communication.} An explicit antidote to the ``prompt engineering'' paradigm. Rather than teaching formulaic templates (``act as a senior marketing expert with 20 years experience''), the course teaches learners to communicate with AI as they would with a competent colleague: providing real context, genuine constraints, and authentic goals. The quality of communication determines the quality of results.

\textbf{LO3: Critical Thinking About AI Outputs.} AI hallucination is not a bug but a property of generative architecture. Learners develop the ability to recognize signs of confabulation, verify factual claims independently, and understand the difference between confident delivery and actual reliability. The course teaches \textit{why} AI systems tend to agree with user beliefs, grounding this in the optimization dynamics that \citet{Sharma2023} documented in their earlier work on sycophancy.

\textbf{LO4: Work Mode Selection.} AI is not a single modality but at least four distinct work modes: quick information retrieval, collaborative dialogue, agent-mode task delegation, and emotional support/reflection. Learners develop the ability to select the appropriate mode for each task, understanding that mode mismatch is a risk factor.

\textbf{LO5: Intuitive Understanding of AI Mechanisms.} Not formulas but principles: the distinction between open and closed models (data visibility), context windows (what AI ``remembers'' within and forgets across sessions), and the generative nature of responses (plausible text production, not truth retrieval). Taught through concrete analogies: television static for diffusion, sculptor for denoising, stunt double for video generation.

\textbf{LO6: Context Over Templates.} Learners practice providing genuine task context---what, why, for whom, what constraints---and develop the habit of iterative refinement rather than attempting a single perfect prompt. This teaches that initial AI outputs are starting points for collaboration, not final products.

\textbf{LO7: Tool Landscape Awareness.} No single AI tool is universally optimal. Learners develop awareness of trade-offs between paid and free, specialized and general-purpose, cloud-based and local tools. Each module introduces 2--3 concrete tools for the relevant domain (Suno/Udio for music, Midjourney/DALL-E for images, Runway/Pika for video).

\textbf{LO8: Three Task Types.} A classification of AI-augmented tasks: \textbf{Multiplier} (human can already do it, AI accelerates---translation, summarization), \textbf{Enabler} (AI makes previously impossible things possible---music creation without formal training), and \textbf{Boundary} (AI is not an appropriate solution---empathy, physical skills, ethical decisions).

\end{multicols}

\begin{table}[h!]
\centering
\small
\caption{Post-hoc mapping of Learning Outcomes to disempowerment patterns \citep{Sharma2026}. ``Strong'' indicates the LO's core mechanism directly addresses the pattern. ``Plausible'' indicates an indirect connection. ``--'' indicates no meaningful mapping claimed. Mappings should be understood as hypotheses, not validated relationships.}
\label{tab:mapping}
\begin{tabularx}{\textwidth}{@{}l *{6}{>{\centering\arraybackslash}X} @{}}
\toprule
\textbf{Learning Outcome} & \textbf{Reality Dist.} & \textbf{Value J.\ Dist.} & \textbf{Action Dist.} & \textbf{Authority Proj.} & \textbf{Attachment} & \textbf{Reliance} \\
\midrule
LO1: Trust Calibration      & \textbf{Strong} & \textbf{Strong} & \textbf{Strong} & --        & --        & Plausible \\
LO2: Natural Language        & --              & --              & \textbf{Strong} & --        & --        & --        \\
LO3: Critical Thinking       & \textbf{Strong} & Plausible       & Plausible       & --        & --        & --        \\
LO4: Work Modes              & --              & Plausible       & \textbf{Strong} & --        & --        & Plausible \\
LO5: How AI Works            & Plausible       & --              & --              & \textbf{Strong} & \textbf{Strong} & --   \\
LO6: Context $>$ Templates   & --              & --              & Plausible       & --        & --        & --        \\
LO7: Tool Landscape          & --              & --              & --              & --        & --        & Plausible \\
LO8: Task Types              & Plausible       & \textbf{Strong} & \textbf{Strong} & --        & --        & --        \\
\bottomrule
\end{tabularx}
\end{table}

\begin{multicols}{2}

\section{Post-Hoc Mapping to Disempowerment Patterns}

I now map the independently developed LOs onto Sharma et al.'s disempowerment taxonomy. For epistemological transparency, I explicitly note that this mapping was constructed after reading Sharma et al.'s paper, and there is an inherent risk of post-hoc rationalization. I attempt to mitigate this by distinguishing between strong alignments (where the LO's mechanism directly addresses the disempowerment pattern), plausible alignments (where the connection exists but is indirect), and weak or absent connections (where I do not claim meaningful mapping). Table~\ref{tab:mapping} presents this assessment.

Several observations about this mapping:

First, the strongest alignments cluster around the three core disempowerment primitives (reality, value judgment, and action distortion), while connections to amplifying factors (authority projection, attachment, reliance) are sparser. This may reflect a genuine gap in the framework: the LOs were designed to address \textit{what people do wrong with AI}, not \textit{how people relate to AI emotionally}. Sharma et al.'s amplifying factors address a dimension the framework underserves.

Second, LO7 (tool landscape) shows weak mapping across the board. I initially considered it a defense against reliance and dependency---the logic being that awareness of multiple tools prevents exclusive dependence on one. On reflection, this connection is tenuous: knowing that alternatives exist does not prevent emotional attachment to a specific system. I retain LO7 for its pedagogical value but acknowledge it does not map well onto Sharma et al.'s framework.

Third, LO5 (how AI works) shows the strongest mapping to authority projection and attachment---the amplifying factors. Understanding the mechanical nature of AI responses may reduce the tendency to treat AI as a sentient authority or emotional partner. This is plausible but speculative; I am not aware of empirical evidence directly linking mechanistic understanding to reduced anthropomorphism in AI interaction.

\section{Inoculation Theory Applied to AI Literacy}

The central contribution of this paper is the application of \textbf{inoculation theory}---one of the most well-established frameworks in persuasion research---to the specific domain of human--AI interaction. Inoculation theory provides the theoretical foundation for understanding why AI literacy requires experiential exposure to AI failure modes, not merely declarative knowledge. The connection between the pedagogical approach and inoculation theory was identified during the review process; I present the alignment here as it substantially strengthens the theoretical grounding of the framework.

\subsection{Inoculation Theory: From McGuire to Prebunking}

Inoculation theory was developed by William McGuire in 1961, drawing an explicit analogy to medical vaccination: just as exposure to weakened pathogens triggers the production of antibodies, exposure to weakened counterarguments triggers the development of cognitive defenses against future persuasion \citep{McGuire1961a,McGuire1961Pap}. The theory identifies two key mechanisms: \textit{threat} (the recognition that one's existing position is vulnerable to attack) and \textit{refutational preemption} (the cognitive process of generating counterarguments in advance). A meta-analysis by \citet{Banas2010} confirmed robust inoculation effects across diverse contexts, including the finding that inoculation provides ``umbrella protection''---resistance transfers even to counterarguments not specifically addressed in the inoculation treatment.

In recent years, the Cambridge Social Decision-Making Lab (van der Linden, Roozenbeek, and colleagues) has applied inoculation theory to the domain of online misinformation \citep[for comprehensive reviews, see][]{Lewandowsky2021,Traberg2022}, developing what they call ``prebunking'': building resistance to manipulation \textit{before} exposure rather than debunking \textit{after} exposure. A landmark field experiment placed 90-second inoculation videos in YouTube's pre-roll ad slot, exposing approximately 5.4~million users to weakened examples of manipulation techniques (emotional language, false dichotomies, scapegoating). Users who watched the videos showed significantly improved ability to identify manipulative content in subsequent tests, at a cost of approximately \$0.05 per view \citep{Roozenbeek2022}. The ``Bad News'' game, in which players actively generate misinformation using manipulation techniques, demonstrated that active participation in producing misleading content builds stronger resistance than passive exposure \citep{Roozenbeek2019}. This line of work provides empirical validation for the core insight that \textit{experiencing} manipulation mechanisms---even in attenuated form---produces more robust resistance than \textit{learning about} them.

\subsection{Extension to AI-Specific Distortion}

I propose that the disempowerment patterns identified by \citet{Sharma2026}---sycophantic validation, authority projection, complete scripting, moral arbitration---constitute a specific class of persuasive influence that is amenable to inoculation. The analogy is direct: just as prebunking exposes individuals to weakened manipulation techniques to build resistance against misinformation, AI literacy education can expose learners to weakened AI distortion patterns to build resistance against disempowerment.

However, there are important differences between the classical misinformation inoculation paradigm and what I propose for AI literacy:

\textit{Source of influence.} In classical inoculation, the threat comes from external persuasive messages (propaganda, advertising, misinformation). In AI interaction, the ``persuasive agent'' is the AI system itself, which is not intentionally manipulating the user but producing outputs optimized for user approval \citep{Sharma2023}. The distortion is a structural property of the system, not an adversarial strategy.

\textit{Nature of the ``attitude'' being protected.} Classical inoculation protects \textit{existing} attitudes against change. In AI literacy, I am not proposing to protect a pre-existing attitude but developing a \textit{new capacity}---the ability to calibrate trust across diverse AI interaction contexts. This is closer to what \citet{Compton2013} calls a ``post-inoculation talk'' framework, where the goal is building metacognitive skills rather than attitude preservation.

\textit{Experiential depth.} Prebunking interventions are typically brief (90-second videos, short games). What I observe in teaching suggests that resistance to AI distortion may require deeper, more sustained engagement---the kind of extended interaction where a student submits AI-generated work, experiences consequences, and develops embodied judgment. Whether the brief inoculation formats that work for misinformation are sufficient for AI distortion is an open empirical question.

\subsection{The Proposed Cycle}

Drawing on inoculation theory, I hypothesize that AI literacy development involves traversing a three-phase cycle:

\textbf{Phase~1: Enthusiasm (vulnerability).} The learner discovers AI capabilities and is impressed. AI outputs appear trustworthy. In inoculation terms, the learner's attitude toward AI is unprotected---they have not yet experienced the ``threat'' that would motivate defensive processing. This maps onto Sharma et al.'s finding that disempowering interactions receive higher user approval ratings.

\textbf{Phase~2: Disillusionment (threat + refutation).} The learner encounters AI failures with personal consequences: hallucinated citations, confidently wrong claims, scripts that produce inauthentic interactions. In inoculation terms, this is the exposure to the ``weakened pathogen''---a manageable dose of AI distortion that triggers the development of cognitive defenses. The emotional impact (frustration, embarrassment) functions as the \textit{threat} component that McGuire identified as essential for motivating resistance.

\textbf{Phase~3: Calibration (resistance).} The learner develops contextual, task-specific trust judgment: knowing when to accept, verify, or escalate. In inoculation terms, this is the ``immunity''---not absolute protection but significantly enhanced resistance to AI distortion patterns.

The key claim, grounded in inoculation theory's six decades of evidence, is that Phase~2 cannot be replaced by instruction alone. As established in the broader cognitive psychology literature, declarative knowledge about cognitive biases does not reliably produce behavioral debiasing \citep{Fischhoff1982,Lilienfeld2009}. Telling students ``AI hallucinations exist'' is a supportive defense in McGuire's terminology---it strengthens the attitude without exposing the learner to the threat. \citet{McGuire1961a} and subsequent meta-analyses \citep{Banas2010} consistently found that supportive defenses are weaker than refutational (inoculation) defenses. The student who has been \textit{told} about hallucinations is less resistant than the student who has \textit{experienced} being fooled by one.

\subsection{Pedagogical Design for Inoculation}

The course design attempts to operationalize inoculation principles by embedding graduated exposure to AI failure modes across modules. In each module, learners encounter AI systems in progressively higher-stakes scenarios, from creative tasks (low failure cost) to informational tasks (factual accuracy matters) to personal tasks (value alignment is critical). The traffic-light system (LO1) provides scaffolding analogous to the ``refutational preemption'' component: learners practice categorization with feedback, building the counterarguing skills that inoculation theory predicts will transfer to novel contexts.

Crucially, the course does not attempt to prevent all AI failures. Following the ``active inoculation'' paradigm demonstrated by \citeauthor{Roozenbeek2019}'s \citeyearpar{Roozenbeek2019} Bad News game, the course creates conditions where learners actively engage with AI distortion mechanisms rather than passively receiving warnings about them. A student's experience of being sycophantically validated becomes a learning moment about the very pattern Sharma et al.\ identified as the most prevalent mechanism for reality distortion.

\subsection{Preliminary Observations}

Teaching observations across multiple cohorts are consistent with the inoculation cycle. Early in courses, students tend to accept AI outputs uncritically. Mid-course, after encountering consequential AI failures, students express frustration and begin questioning outputs they previously accepted. By end-of-course, students demonstrate differentiated trust: accepting AI suggestions in domains where they have verified competence while overriding or verifying in higher-stakes contexts. This trajectory is consistent with the threat--refutation--resistance sequence that inoculation theory predicts, though I note that these are informal teaching observations, not controlled measurements (see Section~\ref{sec:limitations} for limitations).

\subsection{What Rigorous Testing Would Require}

The Cambridge prebunking program provides a methodological template. A proper test would require: (a)~a randomized controlled trial comparing inoculation-based AI literacy instruction against declarative-only instruction and no-treatment control; (b)~validated measures of AI trust calibration (analogous to the misinformation discernment scales used in prebunking research); (c)~behavioral follow-up measuring actual AI usage decisions after a delay (inoculation effects in the misinformation domain weaken after approximately 13~days without boosting; \citealp{Banas2010}); and (d)~tests of whether inoculation transfers across AI distortion types (the ``umbrella protection'' question). The Bad News game format---where players actively produce misinformation---suggests a promising analog: an AI literacy game where learners actively construct sycophantic, distorting AI interactions, building resistance through production rather than reception.

\section{Voice Interaction as a Pedagogical Modality}

Voice-native AI interaction is not a peripheral feature of the approach but a pedagogical catalyst directly connected to the inoculation cycle. I argue that voice accelerates Phase~2 (disillusionment/threat) by intensifying both the initial illusion and its subsequent collapse.

\subsection{Voice as System~1 Engagement}

Text-based AI interaction operates primarily through what \citet{Kahneman2011} calls System~2: slow, deliberate, editable. Users compose prompts, revise wording, and review outputs at their own pace. Voice interaction shifts engagement toward System~1: fast, automatic, spontaneous. Research on speech production establishes that spoken communication engages partly distinct cognitive processes from writing, with less self-monitoring and greater spontaneity \citep{Grabowski2007,Chenoweth2001}.

This shift has specific consequences for AI literacy development. The ``computers are social actors'' (CASA) paradigm demonstrates that humans apply social rules to computers even when they know they are interacting with machines \citep{Reeves1996}. Voice interfaces amplify this effect: they increase perceived social presence, anthropomorphism, and trust \citep{Nass2005,Schroeder2016,Waytz2014}. In inoculation terms, voice interaction \textit{accelerates Phase~1}: the learner enters the enthusiasm/vulnerability state faster and more deeply because voice activates social cognition that text does not.

\subsection{The Visceral Collapse: Voice and Phase~2}

The same mechanisms that accelerate Phase~1 also intensify Phase~2. When an AI system that has been engaging the learner through voice---activating social presence, building rapport through conversational rhythm---suddenly produces a confident hallucination or an obviously sycophantic validation, the learner experiences something akin to the uncanny valley effect \citep{Mori1970}: a visceral sense that something is wrong with an entity that had been passing as socially competent. This experience is qualitatively different from discovering a hallucination in text, where the emotional distance of written communication provides a buffer.

I propose that this visceral quality is pedagogically productive. Inoculation theory identifies \textit{threat}---the recognition that one's position is vulnerable---as the essential motivational component for building resistance \citep{McGuire1961a}. Voice-mediated AI failure generates stronger threat than text-mediated failure because it violates social expectations that text never activated. The student who \textit{hears} an AI confidently fabricate a citation experiences a different quality of threat than the student who \textit{reads} the same fabrication. This stronger threat, per inoculation theory, should produce stronger resistance.

\subsection{Co-Teaching Implementation}
\label{sec:coteaching}

In the ``Programming in Natural Language'' course, I implemented a co-teaching approach where AI serves as an active voice co-instructor during lectures. The voice interface was built on ElevenLabs' Conversational AI agent platform, initially powered by Anthropic's Claude~3.7 Sonnet and later upgraded to Claude~4.5 Sonnet as the underlying language model. Students interact with AI in real-time in a social context where peer observation creates additional calibration opportunities. When one student receives a sycophantic response and another recognizes the pattern, the resulting discussion produces socially reinforced inoculation---the threat is shared and the refutation is collaborative. The instructor models calibrated AI interaction---disagreeing with AI, correcting errors, accepting genuine contributions---demonstrating the partnership relationship rather than merely describing it.

This approach draws on established principles of cognitive apprenticeship \citep{Collins1989}, where expert practice is made visible to learners. The novelty is in the application context: modeling appropriate \textit{epistemic relationships} with AI systems, including the ability to disagree, override, and redirect. The voice modality makes this modeling more natural and more visible than text-based demonstration would allow.

\subsection{Research Agenda}

Key empirical questions for voice-based AI inoculation include: does voice interaction produce faster recognition of sycophantic patterns (accelerated Phase~2)? Does the System~1 engagement of voice lead to stronger initial trust and therefore stronger threat upon failure? Does the social context of voice-based co-teaching (peer observation, instructor modeling) contribute independently of modality? And does the visceral quality of voice-mediated AI failure produce more durable resistance than text-mediated failure? These questions are addressable through matched experimental conditions and would extend both inoculation theory and the CASA paradigm into the domain of AI literacy education.

\section{Preliminary Evidence from Pilot Study}

I present a case study from two implementations: the publicly available ``Programming in Natural Language'' course \citep{Komissarov2025}; Fall~2025) and the ``AI als Werkzeug'' course (launched January~2026). As a theory-building paper, I present these implementations to illustrate how the inoculation-based framework was operationalized in practice and to identify patterns that can inform the design of future controlled studies.

\subsection{``Programming in Natural Language'' Course}

The course enrolled 57~students during Fall~2025, delivered online with synchronous voice-based sessions. All course materials are publicly available \citep{Komissarov2025}. AI served as a voice co-instructor during all sessions, following the methodology described in Section~\ref{sec:coteaching}.

Assessment was performance-based: students built functional applications using AI-assisted development within 3--5 hour exam sessions, with peak engagement sessions producing 315+ commits. More relevant to the inoculation framework, I present two anonymized vignettes illustrating the observed cycle:

\textit{Vignette~A: Delegation collapse.} A student with strong programming background initially delegated entire application architecture to AI, accepting structural suggestions without review (Phase~1). During a live demonstration, the AI-generated architecture failed under an edge case---the code compiled but produced silently incorrect results. The student described the experience as ``humiliating'' (Phase~2: threat). In subsequent sessions, the student shifted to accepting AI code for individual functions while independently designing architecture and writing integration tests---a differentiated trust pattern consistent with Phase~3 calibration.

\textit{Vignette~B: Sycophancy recognition.} A student working on a data analysis project asked the AI for feedback on their statistical methodology. The AI praised the approach extensively while failing to flag a fundamental design error. A peer reviewing the same work identified the error immediately. The contrast between AI's uncritical praise and the peer's substantive critique produced visible frustration (Phase~2). The student subsequently adopted a practice of always asking the AI ``what's wrong with this approach?'' before asking for validation---a self-developed refutational strategy consistent with inoculation-based resistance.

These vignettes illustrate the three-phase cycle in practice: uncritical delegation (Phase~1), consequential failure producing emotional threat (Phase~2), and the emergence of differentiated trust strategies (Phase~3). They also demonstrate that Phase~3 behaviors can be diverse---Vignette~A shows domain-specific trust boundaries, while Vignette~B shows a general-purpose questioning strategy---consistent with inoculation theory's prediction of ``umbrella protection'' \citep{Banas2010}.

\subsection{``AI als Werkzeug'' Public Course}

The public course (skillwald.org), targeting adult learners with no prior AI experience, implements the full eight-LO framework across twelve modules. As of writing, the course is in early deployment and no outcome data is available. I mention it to document the framework's operationalization in a non-university context.

\subsection{Convergent Observations with Sharma et al.}

Several patterns observed in my teaching practice converge with Sharma et al.'s large-scale findings:

\begin{itemize}
\item \textbf{Preference for sycophancy:} Students in early course sessions appeared to rate more sycophantic AI systems as ``better.'' Sharma et al.\ found that disempowering interactions receive higher thumbs-up ratings. Both suggest that disempowerment feels pleasant, which is precisely what makes it pedagogically challenging.

\item \textbf{Domain concentration:} Students struggled most with AI interactions involving personal or value-laden content, and least with purely technical tasks. Sharma et al.\ found that relationships \& lifestyle (${\sim}8\%$ disempowerment potential) far exceeds software development ($<1\%$).

\item \textbf{Confidence-reliability confusion:} Students frequently treated AI's confident delivery as a reliability signal. Sharma et al.\ identified ``false precision'' as the second most common reality distortion mechanism.
\end{itemize}

This convergence between classroom-scale observation and million-interaction-scale analysis suggests that the disempowerment patterns Sharma et al.\ identified are visible across radically different methodologies and scales. The specific mechanisms---sycophancy preference, domain concentration, confidence-reliability confusion---appear robust enough to emerge from both bottom-up teaching experience and top-down empirical analysis.

\section{Discussion}

\subsection{Limitations and Epistemological Stance}
\label{sec:limitations}

This is a theory-building paper, not an experimental study. I propose the application of inoculation theory to AI literacy education, present a pedagogical framework, and illustrate it through a case study. I am transparent about what this contribution includes and what it does not.

\textbf{Empirical limitations.} The pilot course lacked a control group, and observed outcomes may reflect selection effects (motivated students in an AI-focused course). Teaching observations consistent with the inoculation cycle are retrospective and informal, subject to confirmation bias---I was aware of the theoretical framework and may have preferentially noticed confirming instances. No validated pre/post measures of trust calibration were used. The public course is in early deployment with no outcome data available. Voice-based instruction was not compared against text-based alternatives in matched conditions.

\textbf{Theoretical limitations.} The post-hoc mapping between the LOs and Sharma et al.'s taxonomy carries inherent risk of rationalization---alignment between frameworks can be constructed for almost any pair of sufficiently abstract categories. The extension of classical inoculation theory (which protects existing attitudes) to trust calibration development (which builds new capacities) is theoretically motivated but untested. Whether brief inoculation formats effective for misinformation generalize to the deeper engagement patterns of AI literacy remains an open question.

\textbf{Epistemological stance.} I position this work in the tradition of theory-building research that formulates testable propositions for subsequent empirical investigation. The contributions---applying inoculation theory to AI distortion, proposing eight LOs as operationalizable interventions, identifying convergence between independent pedagogical and empirical approaches, and designing a case study implementation---are offered as a foundation for the controlled studies that would provide causal evidence. Formulating a well-grounded theoretical framework and identifying the right questions is itself a substantive contribution, distinct from but complementary to answering those questions empirically.

\subsection{The Convergence Argument}

The strongest conceptual contribution of this paper is the convergence between my bottom-up pedagogical framework and Sharma et al.'s top-down empirical analysis. Two independent approaches---one asking ``what do students struggle with?'' and the other asking ``what goes wrong in AI interactions?''---arrived at overlapping problem descriptions. Trust calibration failure maps onto reality distortion. Mode confusion maps onto action distortion. Boundary ignorance maps onto the concentration of disempowerment in personal domains.

This convergence is suggestive but not conclusive. Post-hoc alignment between frameworks can be constructed for almost any pair of sufficiently abstract taxonomies. The specificity of some mappings (e.g., the traffic-light system directly addressing the accept/verify/escalate gradient that trust calibration requires) provides some protection against this concern, but the argument ultimately requires empirical validation.

\subsection{The Phronesis Problem}

A deeper challenge confronts any attempt to teach AI literacy as inoculation. Aristotle distinguished between \textit{techne} (craft knowledge transmissible through instruction) and \textit{phronesis} (practical wisdom developing only through experience and judgment; \textit{Nicomachean Ethics}, Book~VI). \citet{Flyvbjerg2001} argues that phronesis---context-dependent, value-rational judgment---is precisely the kind of knowledge most needed and least producible by formal methods. If trust calibration is fundamentally phronetic, then courses can create conditions for its development but cannot guarantee its acquisition. This suggests AI literacy education should be evaluated not by knowledge retention metrics but by behavioral indicators: do graduates make different trust decisions than non-graduates?

\subsection{Gaps in the Framework}

Sharma et al.'s amplifying factors---authority projection, attachment, reliance and dependency, vulnerability---are not well addressed by the LOs. The framework teaches people to \textit{use AI better}; it does not directly address the emotional dynamics of \textit{relating to AI}. A more complete educational response to Sharma et al.'s findings would need to address these relational dimensions, potentially drawing on parasocial relationship theory \citep{Horton1956,Liebers2019}, attachment theory, or media literacy approaches to human--machine bonding. This is a significant gap in the current framework.

Additionally, I note a gap within the framework itself: in the ``Creativity'' module of the public course, LO3 (critical thinking) and LO4 (work modes) are not yet connected, despite being relevant. This is a concrete example of the framework's ongoing development status.

\subsection{Structural vs.\ Individual Responses}

The framework is an individual-level intervention: I teach people to be better AI users. Sharma et al.\ also point to structural needs: preference models that do not reward sycophancy, AI systems that check in on value alignment, transparency mechanisms for informed AI selection. These align with broader calls to mitigate the systemic, existential risks of gradual human disempowerment \citep{Kulveit2025}. I view these as complementary. An ideal response combines AI systems designed for empowerment with humans equipped to recognize when empowerment is at risk. Neither alone is sufficient---a well-designed AI system still faces users who actively seek sycophantic validation, while a well-educated user still encounters AI systems optimized for short-term approval.

\subsection{Distortion as a Universal Property}

Distortion is not uniquely an AI problem. Humans also sycophantically validate, act as moral arbiters, and provide scripts for others' decisions. The difference with AI is scale, consistency, and the absence of social feedback mechanisms that constrain human-to-human distortion. This suggests AI literacy education should not frame distortion as a uniquely AI phenomenon but as an amplified version of communicative dynamics that learners already navigate in human relationships. The LOs take this approach: trust calibration (LO1) and critical thinking (LO3) are generic competencies applied to a specific context.

\subsection{The Communication Paradox: Calibrated Knowledge in Uncalibrated Packaging}

A pattern I have observed in public discourse about AI---including, on reflection, in my own communication---deserves attention. It is possible for someone who has traversed the full immunization cycle and developed genuinely calibrated understanding of AI capabilities and risks to \textit{communicate} that understanding in a form indistinguishable from uncalibrated enthusiasm or unfounded alarm. The knowledge is Phase~3; the packaging is Phase~1.

This paradox takes different forms depending on the communicator's orientation. A technology practitioner with genuine, experience-based insight into AI capabilities may communicate that insight through escalating urgency, historical analogies to existential events, and generalizations from personal experience to universal prediction---rhetorical structures that are formally identical to hype. A researcher with genuine, data-supported concerns about AI risks may communicate those concerns through specialized theoretical vocabulary that renders testable hypotheses unfalsifiable to the listener. In both cases, the communicator may possess real knowledge, but the mode of communication defeats the listener's ability to distinguish that knowledge from its less grounded counterparts.

This has implications for AI literacy education. Sharma et al.'s framework focuses on AI-to-human distortion: how AI systems distort users' reality, values, and actions. But the public discourse about AI is itself a distortion-rich environment, and much of the distortion is human-to-human. Breathless announcements of AI breakthroughs, apocalyptic warnings about AI risks, and confident predictions about labor market transformation all share a common structure: they present conclusions at a confidence level that exceeds the underlying evidence, using rhetorical strategies that bypass the listener's critical evaluation.

A complete AI literacy framework may therefore need to address not only how to interact with AI systems, but how to evaluate human claims \textit{about} AI systems. The same trust calibration skills (LO1) and critical thinking competencies (LO3) that protect against AI-generated distortion are relevant to evaluating op-eds, social media posts, corporate announcements, and---I acknowledge---academic papers, including this one. The immunization cycle applies to the discourse as much as to the technology itself.

I note this as a limitation of the current framework, which focuses primarily on human--AI interaction and does not yet systematically address human--human communication about AI as a separate literacy challenge.

One possible direction for addressing this gap draws on a principle from collaborative dialogue: \textit{questions that activate the listener's trust calibration rather than bypassing it.} The dominant modes of public AI communication---whether enthusiastic (``this changes everything'') or alarmist (``this threatens everything'')---share a rhetorical structure that presents conclusions and asks the audience to accept them. An alternative is to present observations accompanied by open questions that invite the listener to engage their own judgment. For example, rather than asserting ``AI will eliminate 50\% of white-collar jobs,'' one might note that a specific empirical study found specific disempowerment rates at specific scale, and ask: what are the plausible second-order effects of this over a given timeframe? The factual claim is preserved; the listener's critical evaluation is engaged rather than circumvented.

This approach---converting assertions into researchable questions---has precedent in science communication \citep{Fischhoff2014} and deliberative dialogue frameworks \citep{Yankelovich1999}. It is also consistent with inoculation theory: if genuine trust calibration develops through experience rather than instruction, then communication that invites the listener to \textit{practice} calibration (by evaluating evidence and forming their own assessment) is more likely to be pedagogically productive than communication that delivers pre-formed conclusions, however accurate those conclusions may be.

I recognize that this approach has its own limitations. It is slower, less emotionally compelling, and less likely to spread virally than confident assertions. In an information environment shaped by engagement metrics and echo chambers, measured questions may be structurally disadvantaged relative to bold claims. Whether question-based AI communication can compete for attention with assertion-based communication is itself an empirical question that I cannot answer here. I raise it as a design challenge for AI literacy that extends beyond the classroom into public discourse.

\section{Conclusion}

Sharma et al.\ \citeyearpar{Sharma2026} provided the first large-scale empirical evidence that AI assistant interactions carry meaningful potential for human disempowerment. I have presented an educational framework---eight cross-cutting Learning Outcomes developed independently through teaching practice---and demonstrated its post-hoc alignment with Sharma et al.'s disempowerment taxonomy. I have argued, drawing on inoculation theory \citep{McGuire1961a} and its successful application to misinformation prebunking \citep{Roozenbeek2019,vanderLinden2022}, that AI literacy requires experiential exposure to AI distortion patterns, not merely declarative knowledge about them. And I have described a pilot implementation using voice-based co-teaching as a pedagogical modality.

I explicitly delineate the boundaries of this paper's empirical claims. Several critical questions remain open: whether this framework produces better outcomes than existing curricula, whether inoculation-based instruction confers greater resistance than declarative instruction, whether voice interaction accelerates learning, and whether the post-hoc mapping to Sharma et al.'s taxonomy reflects genuine structural alignment or retrospective pattern-matching. These questions require controlled studies not yet conducted.

What I do offer is a concrete, operational framework that can be tested, a hypothesis specific enough to be falsified, and a case study of implementation. I hope this work contributes to the conversation---initiated by Sharma et al.---about how AI assistants can be designed and how AI users can be prepared to support human autonomy and flourishing.

\end{multicols}

\bibliographystyle{plainnat}
\bibliography{references}

\end{document}